\documentclass[aps,prb, showpacs, twocolumn, superscriptaddress]{revtex4-1}
\usepackage[utf8]{inputenc}
\usepackage{graphicx}
\usepackage{pstricks}
\usepackage{array}
\usepackage{natbib}
\usepackage{amsmath}
\usepackage{pifont}
\usepackage{dsfont}
\usepackage{multirow}
\usepackage{amssymb}
\usepackage{wasysym}

\def\be{\begin{equation}}
\def\ee{\end{equation}}
\def\tb{\textbf}

\begin{document}

\title{Dynamical control of interference using voltage pulses in the quantum regime}

\newcommand{\spsms}{CEA-INAC/UJF Grenoble 1, SPSMS UMR-E 9001, Grenoble 
F-38054, France}

\author{Benoit Gaury}
\affiliation{\spsms}
\author{Xavier Waintal}
\affiliation{\spsms}
\date{\today}

\begin{abstract}
As a general trend, nanoelectronics experiments are shifting toward frequencies
so high that they become comparable to the device's internal characteristic time
scales, resulting in new opportunities for studying the dynamical aspects of
quantum mechanics.  Here we theoretically study how a voltage pulse (in the
quantum regime) propagates through an electronic interferometer (Fabry-Perot or
Mach-Zehnder). We show that
extremely fast pulses provide a conceptually new tool for manipulating quantum information: 
the possibility to dynamically engineer the interference pattern of a quantum
system.  Striking physical signatures are associated with this new regime: restoration of the
interference in presence of large bias voltages; negative currents
with respect to the direction of propagation of the voltage pulse; and oscillation
of the total transmitted charge with the total number of injected electrons.
The present findings have been made possible by the recent unlocking of our capability for simulating time-resolved quantum nanoelectronics of large systems. 
\end{abstract}

\maketitle

\noindent The quantum dynamics of discrete levels is by now so well understood that systems of
several qubits (and photonic modes) are routinely engineered and addressed using microwave signals.
In contrast, very little is known about the continuum, i.e. the dynamics of
degrees of freedom which are allowed to propagate inside a system. A few works
propose setups for ``flying qubits"\cite{Bertoni2000,
Buttiker2004,BeenakkerFQ,Degiovanni,flying_qbit2,Lebedev2005} that encode the quantum
information into the paths taken by the electrons. Those systems could be
realized in Mach-Zehnder interferometers in the quantum Hall regime
\cite{MachZender_Heiblum,20mum_20mK,Haack} or Aharonov-Bohm geometries
\cite{flying_qbit}. 
Before designing any of those circuits however, a necessary step is the understanding of the basic, potentially new, physics
associated with the time resolved dynamics in delocalized nanoelectronic systems. 

Two competing kinds of dynamical excitations have emerged to inject electrons in nanoelectronic devices. In the first, one fills up
the state of a small quantum dot  and then rapidly increases its
energy to release the electron inside the system\cite{Single_e_source}. This setup allows the electrons to be injected
one by one with a rather well defined energy, but badly defined
releasing time. In the second -- on which we shall focus -- one simply uses an Ohmic contact to apply a
voltage pulse $V(t)$ to the device (well defined in time but ill defined in energy). In a single mode device, such a voltage
creates a current $I(t)=(e^2/h) V(t)$ which injects 
\begin{equation}
\bar n = \int dt \frac{eV(t)}{h}
\end{equation}
electrons inside the system. A voltage pulse will be said to be in the quantum regime when roughly $\bar n\approx 1$ electron is injected and the
electronic temperature is smaller than the energy scales associated with the height $V_P$ and duration $\tau_P$ of the pulse ($\hbar/\tau_P$).

In a series of seminal works, Levitov {\it et al.} studied the properties of pulses of Lorentzian shape\cite{Lesovik1994,Levitov1996, Lorentzian_pulses,Ivanov}. While they found a featureless time-dependent current, they predicted that, in contrast, the {\it current noise} could oscillate with the amplitude of the pulse, with the possibility to build noiseless quantum excitations for the particular Lorentzian shape. Recent experiments are beginning to address these proposals\cite{Gabelli,McEuen2008,Glattli}. In particular, the quantum regime was reported recently in the group of D.C. Glattli\cite{Glattli}.
Here we report on the new non-trivial physics that emerges when those voltage pulses are used to inject charge excitations in an electronic interferometer. We find that ultra-fast pulses permit the  {\it dynamical control} of the relative phases of the different paths taken by the electrons, therefore providing  means to dynamically engineer the coherent superposition of the traveling waves. We first focus on a simple Fabry-Perot interferometer in one dimension followed by full-scale simulations of a two dimensional Mach-Zehnder interferometer in the quantum Hall regime.

\begin{figure*} 
    \includegraphics[width=\textwidth]{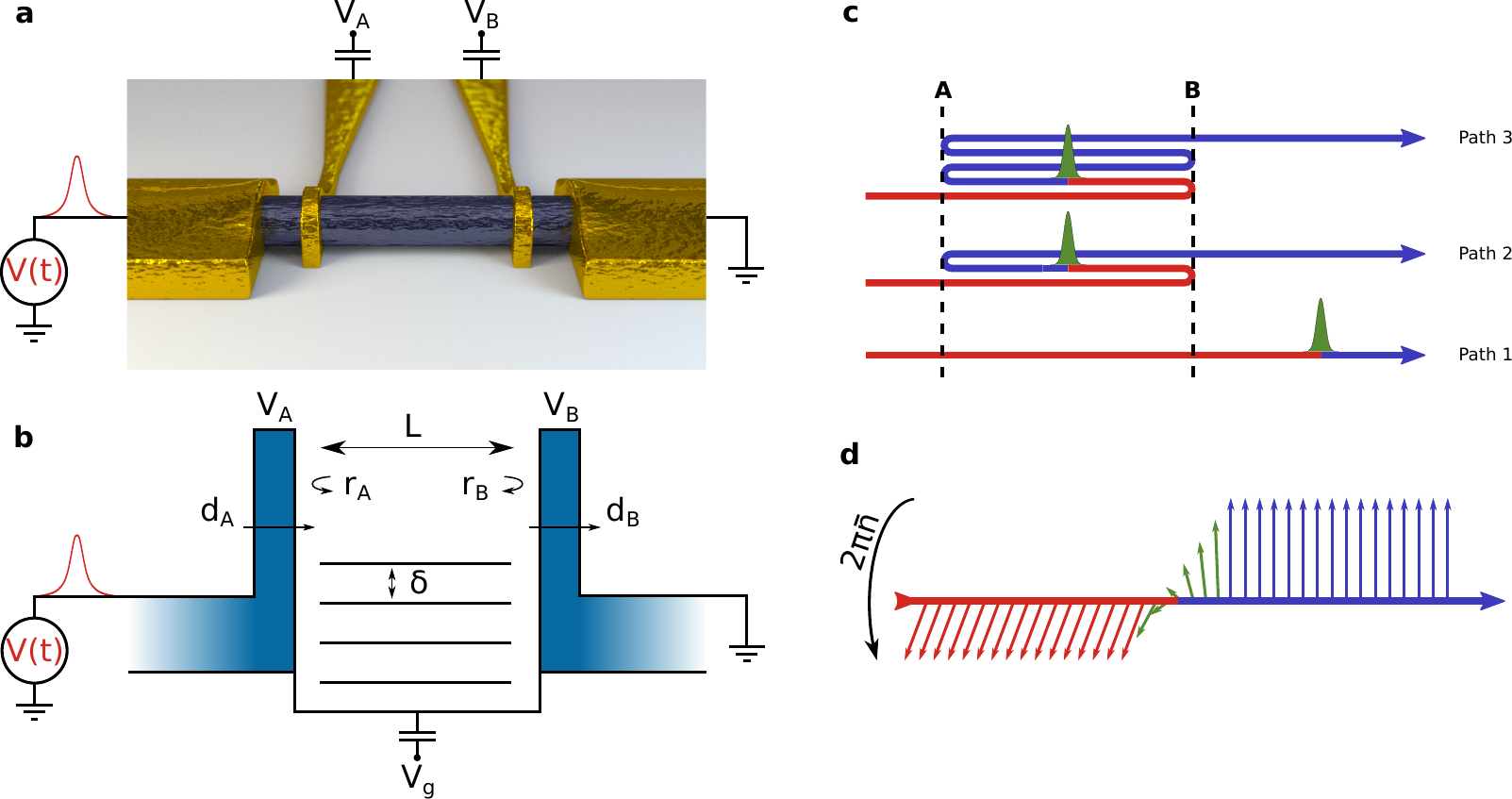} 
    \caption{\tb{Schematic of the Fabry-Perot cavity and of the main physical
mechanism.} (\tb{a},\tb{b}) Schematic of our setup, a quantum wire
connected to two electrodes.  Two barriers A and B separated by a distance $L$
are placed along the wire
    and a Gaussian voltage pulse $V(t)$ is sent from the left. The barriers are characterized by the barrier heights
    ($V_{A}$ and $V_{B}$) or equivalently by their reflection and
    transmission amplitudes denoted respectively $r_{A}, r_{B}$ and $d_{A},
    d_{B}$.  A gate voltage $V_{g}$ allows one to shift the position of the
    resonant levels of the cavity. The mean level spacing between the discrete levels of the cavity is $\delta=h/(2\tau_F)$ where $\tau_F$
    is the ballistic time of flight from A to B.
    (\tb{c}) Schematic of the physical mechanism for the dynamical control of the interference: as the pulse propagates along the different trajectories, a phase difference $2\pi \bar n$  appears between the front (blue) and the rear (red) resulting in a modification of the interference pattern.
    (\tb{d}) Graphical representation of Eq.~(\ref{dw}) that gives the structure of a voltage pulse in terms of a ‘‘phase domain wall".
    \label{fig:fp_model} } 
\end{figure*}

\bigskip
\noindent \tb{Results}\newline
\noindent \tb{Fabry-Perot cavity.}
Fig.~\ref{fig:fp_model}a and \ref{fig:fp_model}b show our model Fabry-Perot system: it consists of a quantum wire connected to two metallic electrodes. The quantum wire is made into a Fabry-Perot interferometer by means of two barriers (A and B) which can be defects in the wire, gates (as in the sketch) or simply the Schottky barriers that naturally form at the
wire-electrode interfaces. Such Fabry-Perot interferometers are standard
devices of nanoelectronics and their DC properties have been extensively
measured \cite{FabryPerot_pioneer,FabryPerot_carbon_nano,InAsFP}.
The basic properties of this interferometer can be
understood within an elementary theory.  Each barrier A (and B) is described by
the amplitude of probability $d_A$ ($r_A$) for an incident electron to be
transmitted (reflected). Summing up the probability amplitudes for all the trajectories (direct transmission: $d_B d_A$, one back and forth bouncing: $d_B r_A r_B d_A$...), the total amplitude of probability for an electron to be transmitted reads,
\be 
\label{qu} 
d_{AB}(E) = \frac{d_{A} d_{B}}{1-r_{A}r_{B} z}.  
\ee 
The factor
$z$ corresponds to the phase $z=e^{i2kL}$ accumulated by the electron during the
time between two collisions ($L$ distance between the scatterers, $k$ electron
momentum).  $z$ can also be rewritten as $z=e^{i2\tau_F E/\hbar}$, where $\tau_F$ is the
time of flight between A and B, and $E$ is the incident energy (our analytical treatment ignores the small
energy dependence of $\tau_F$, $d_A$,$d_B$\dots but our numerics fully accounts for it). When $E$ is at resonance
with the eigenenergies $E_n=n\delta + eV_g$ of the cavity formed by A and B ($\delta=h/(2\tau_F)$: mean level spacing, $eV_g$: potential shift due to a nearby electrostatic gate),  $d_{AB}$ shows a sharp peak and reaches perfect transmission. 

Let us now apply a voltage pulse $V(t)$ of duration $\tau_P$ and maximum intensity $V_P$ to the left electrode. Defining the transmitted current $I_t(t)$ just after the second barrier, the observable of interest to us will be the total number $n_t=\int I_t(t) dt/e$ of electrons transmitted through the system. $n_t$ can be directly measured experimentally and requires much less effort than e.g. noise measurements. In an actual experiment, one would measure the {\it DC} current $I_{dc}$ upon sending periodic trains of pulses through the system. Indeed, by periodically applying the above pulse with a period $\Theta \gg \tau_P$, one simply finds $I_{dc}=en_t/\Theta$.

\begin{figure*}
    \includegraphics[width=\textwidth]{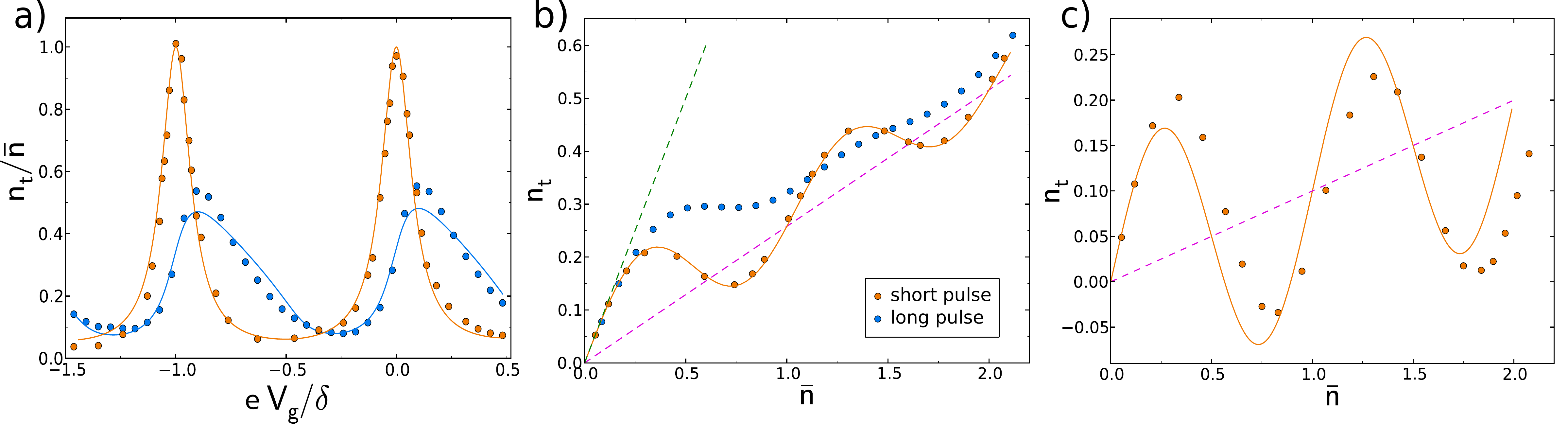}
    \caption{\tb{Total charge transmitted $n_t$.} (\tb{a})
$n_t$ as a function of gate voltage $eV_g/\delta$. (\tb{b},\tb{c}) $n_t$
as a function of total injected charge $\bar n$.  The symbols correspond to
numerical data for short (blue, $\tau_P=\tau_F/7$) and long  (red,
$\tau_P=3\tau_F$) pulses while the full lines correspond to the analytical
results for $\tau_P\ll\tau_F$ (blue) and $\tau_P\gg\tau_F$ (red). (a):
$V_P=0.5\delta$ and $D_A = D_B = 0.5$. (\tb{b},\tb{c}) System at
resonance and $V_P$ is varied with $D_A = D_B = 0.5$ (\tb{b}), and $D_A = D_B =
0.1$ (\tb{c}).  Dashed lines: $n_t=D_{AB}\bar n$ (green) and
$n_t=D^{Cl}_{AB}\bar n$ (magenta).  \label{fig:gp_sine} } 
   
\end{figure*}

The limit of long pulses $\tau_P \gg \tau_F$ is rather trivial: as $V(t)$ varies very slowly, at each instant the current follows the DC I-V characteristics of the system: $I_t(t)= I_{dc}[V(t)]$
(adiabatic limit) which can be obtained from the Landauer formula. In this limit, $V_P\ll \delta$ (linear regime) leads to $n_t=  |d_{AB}(E_F) |^2 \bar n$ ($E_F$: Fermi level), while for large voltages $V_P\gg \delta$ (classical limit) the interference pattern is washed out and one obtains $n_t=  D_{AB}^{cl} \bar n$, where the {\it
classical} (or incoherent) probability $D_{AB}^{cl}$ corresponds to the addition
law of the {\it probabilities} associated with the different paths~\cite{Datta},
\be 
\label{cl} 
D_{AB}^{cl} = \frac{D_{A}D_{B}}{1-R_{A}R_{B}} 
\ee 
(capital
letters $D$ or $R$ correspond to the probabilities associated with the respective
amplitudes so that $D_A=|d_A|^2$). Eq.~(\ref{cl}) is essentially identical to
Eq.~(\ref{qu}) upon replacing  amplitudes by probabilities. So far, we have made rather standard
predictions which are easily reproduced by our numerical simulations:
the blue symbols in Fig.~\ref{fig:gp_sine} show that $n_t$ oscillates with the gate voltage $V_g$ (Fig.~\ref{fig:gp_sine}a)
and increases monotonously with $V_P$ (Fig.~\ref{fig:gp_sine}b). Fig.~\ref{fig:gp_sine}a has been
calculated with an intermediate value of $V_P\approx 0.5\delta$ so that the
contrast of the interference pattern is not very large. 

\begin{figure*}
    \includegraphics[width=\textwidth]{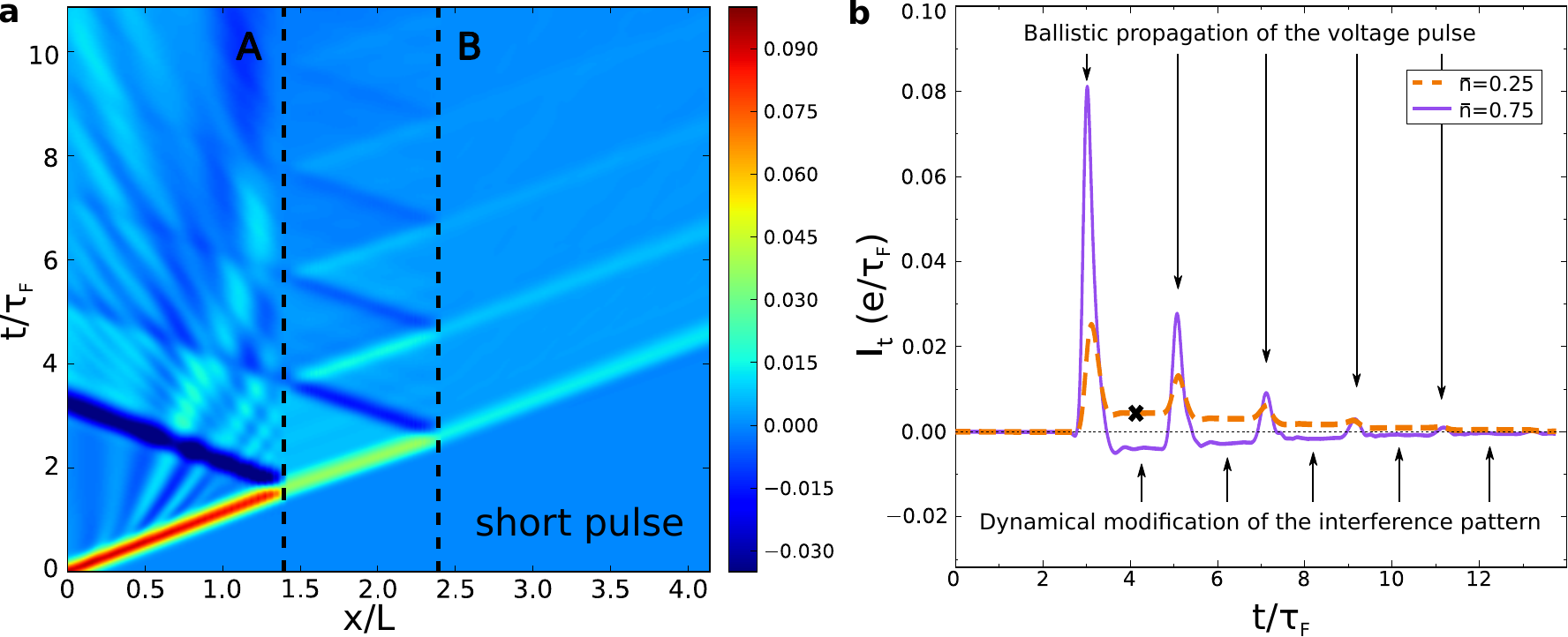} 
    \caption{\tb{Propagation of the voltage pulse.} (\tb{a}) Local current $I_t(x,t)$ as a function of space (in unit of the length $L$ of the cavity) and time (in unit of $\tau_F$)
    for $V_P=1.5\delta$, $\tau_P=\tau_F/3.5$ and the cavity is at resonance. The dashed lines
    indicate the positions of the barriers. 
    (\tb{b}) $I_t(x_0,t)$ for $x_0=2.5 L$ on the right of the second barrier $B$.
    In orange: $V_P=1.5\delta$, in purple: $V_P=4.5\delta$ and
    $\tau_P=\tau_F/3.5$. The black cross ($\boldsymbol{\times}$) marks
    the time associated with Fig.~\ref{fig:fp_model}c.
\label{fig:xt-cards_currents} }
   
\end{figure*}

Having established the adiabatic limit, we can now turn to the more interesting limit of short
pulses $\tau_P\ll\tau_F$ for which a proper time-resolved quantum theory is
compulsory.  Let us make a naive guess: a very short pulse can be viewed as a
very localized  perturbation that will propagate ballistically through the wire.
Monitoring the current after the barriers, one observes a narrow peak
when the perturbation has propagated up to the observation point. $2\tau_F$
later one observes a second peak corresponding to trajectories with one
reflection on each barrier, new peaks (of increasingly
smaller amplitudes) arrive sequentially every $2\tau_F$. As the perturbations coming
from different trajectories do not coincide in time, they cannot interfere and
one expects to
observe the ‘‘classical" addition law $n_t=  D_{AB}^{cl} \bar n$. The argument
can also be made in the energy domain: a fast pulse excites electrons to a large
spread in energy which results in an effectively random phase $z$ and the
interference pattern gets washed out. A rapid glance at the numerics does indeed
confirm this picture: Fig.~\ref{fig:xt-cards_currents}b shows the monitored
current  $I_t(t)$  which clearly shows the peaks described
above. Perhaps more transparent is the corresponding color map of the local current
 $I(x,t)$ (Fig.~\ref{fig:xt-cards_currents}a)  where the different trajectories with multiple
reflections are clearly visible. In contrast, long pulses (not shown) have an essentially featureless 
current $I_t(t)$ of the same shape as the voltage pulse. 

The story could end here: slow pulses allow one to observe the interference effects (wave aspect of
quantum mechanics) while fast pulses give access to the ballistic propagation and
reflection/transmission of the charges injected by the pulse (particle aspect of quantum mechanics).
A  deeper look at the numerics reveals however a handful of rather counter-intuitive physical
effects. First, one observes in the $I_t(t)$ plot of Fig.~\ref{fig:xt-cards_currents}b
that the current does not vanish in between consecutive peaks.
Second, Fig.~\ref{fig:gp_sine} shows that the total number of transmitted electrons in
fact oscillates strongly with the gate voltage (Fig.~\ref{fig:gp_sine}a) in total
contradiction with the above picture. 
Indeed, upon using faster pulses, one
actually restores the interference pattern that was somewhat smeared in the long
pulse case. Third, and even more striking, are Fig.~\ref{fig:gp_sine}b and
Fig.~\ref{fig:gp_sine}c which show that the number of transmitted electrons
actually {\it oscillates } with the number of injected electrons $\bar n$.
Fig.~\ref{fig:gp_sine}c is particularly intriguing: for $\bar n=0.8$, $n_t$,
e.g. the DC current for a train of pulse, is {\it negative}. In other word,
one raises the energy of the electrons on the left
and the electrons flow {\it toward} the left electrode.

\bigskip
\noindent \tb{Dynamical control of the interference pattern.}
To understand this regime of fast pulses, one needs to develop a proper
representation of what a fast voltage pulse really does to the electronic
wave function. The naive image where a voltage pulse generates some sort of localized 
wave packet that propagates through the system is, to a large extent, wrong.
In contrast, stationary delocalized waves already exist before the pulse. Ignoring for a moment the presence of the interferometer (barriers), the stationary wave function is a simple plane wave $\Psi(x,t)=e^{ikx-iEt}$. Upon
applying a voltage pulse $V(t) \theta(-x)$ (we suppose that the voltage drop is very abrupt spatially for the sake of the argument, $\theta (x)$ is the Heaviside function), the energy of the wave is increased and the wave function starts to accumulate an extra phase $\phi(t) = \int_{-\infty}^{t} du\ eV(u)/\hbar$ for $x<0$. Noting that $\lim_{t\rightarrow\infty}\phi(t)=2\pi \bar n$, one finds that the wave function after the pulse takes the form
\begin{align}
\Psi(x,t)= e^{-i2\pi \bar n+ikx-iEt/\hbar} \ \ &{\rm for} \ \ x<0 \nonumber \\
\Psi(x,t)= e^{+ikx-iEt/\hbar} \ \ &{\rm for} \ \ x>0.
\end{align} 
The effect of a voltage pulse is therefore to generate a
kink in the {\it phase} of the electronic wave function $\Psi(x,t)$ (see Fig.~\ref{fig:fp_model}d for a schematic).
In other words, what propagates is
essentially a ‘‘phase" domain wall between two regions which are characterized by an $e^{i2\pi \bar n}$ phase difference. Phases in quantum mechanics cannot be observed directly and one
has to resort to interferences between different paths to observe them. The role of the electronic interferometers
used in this study is to introduce these different paths. While the argument above is very naive, it correctly captures the main feature of the wave function which reads
(for a linearized spectrum),
\be
\label{dw}
\Psi(x,t)= e^{-i\phi(t-x/v)+ikx-iEt/\hbar} 
\ee 
where $v=(1/\hbar) \partial E/\partial k$ is the group velocity. 
 
Let us now return to our Fabry-Perot cavity. In this case, the stationary wave
is not a simple plane wave but a superposition of several waves corresponding
to the different paths that the electrons can take (with zero, one, two...
reflections) as shown in  Fig.~\ref{fig:fp_model}c. When a voltage pulse is sent
through this superposition of paths, it propagates through the various paths.
Fig.~\ref{fig:fp_model}c corresponds to a snapshot at a particular time where
the pulse has emerged from the direct path (path 1 of stationary amplitude
$d_B d_A$) but not yet from the longer trajectories with multiple reflections
(path 2 of amplitude $d_B (r_A r_B) d_A$, path 3 of amplitude $d_B (r_A r_B)^2
d_A$\dots). The time at which this snapshot is taken corresponds to the cross
in the $I_t(t)$ plot of Fig.~\ref{fig:xt-cards_currents}b. If one looks at the
wave function just after the barrier B at that particular time, one finds that
the amplitude of path 1 has an extra phase $e^{i2\pi\bar n}$ compared with its
stationary value
(rear of the pulse as compared with paths 2, 3,... that are still in the front of the pulse). Therefore at this
particular time, the total amplitude is $e^{i2\pi\bar n} d_B d_A + d_B (r_A r_B) d_A + d_B (r_A r_B)^2 d_A$\dots
and is {\it dynamically modified } with respect to its stationary value. As time increases, the pulse will
emerge from path 2, path 3...and the factor $e^{i2\pi\bar n}$ will progressively spreads to all trajectories until
one recovers the stationary amplitude (up to a now irrelevant global
$e^{i2\pi\bar n}$ phase factor). The above argument explains the origin of the
plateaus observed in Fig.~\ref{fig:xt-cards_currents}b. The value of the current
at these plateaus obviously oscillates with $2\pi \bar n$ which consequently
explains the oscillations of $n_t$.
This mechanism, to which we refer to as the {\it dynamical control of the interference pattern}, is the main new concept
of this paper.

In order to make the above argument quantitative, and in particular properly take into account the Fermi statistics
for the filling of the stationary states, we perform the analytical calculation of $n_t$ with a ``photo-assisted tunneling" formula (Eq.~(95) of ref.~\onlinecite{Twave_formalism}),  
\begin{align}
    \label{srt} 
    n_{t} = &\int \frac{dE}{2\pi}\frac{dE'}{2\pi} |d_v(E'-E)d_{AB}(E')|^2
    [f(E)-f(E')] 
\end{align} 
where $d_v(E'-E)$ is the amplitude of probability for
an electron with energy $E$ to be transferred to the energy $E'$ by the pulse and $f(E)$ the Fermi function. 
$d_v(E'-E)$ is essentially the Fourier transform of the phase $e^{-i\phi (t)}$. The calculation of
$n_t$ for fast pulses yields (see the Methods section),
\begin{align} \label{wp} n_{t} &= D_{AB}^{cl}\ \bar{n} + [D_{AB}(V_g) -
D_{AB}^{cl}] \frac{\sin(2\pi\bar{n})}{2\pi}\nonumber\\ &-
\frac{2D_{AB}(V_g)D_{AB}^{cl}r_{A}r_{B}}{\pi D_{A}D_{B}} \sin^{2}(\pi \bar{n})
\sin(2\pi V_{g}/\delta) .
\end{align} 
Equation~(\ref{wp}) contains two contributions of different kind: the first term, ``particle" like, accounts for the
ballistic propagation of the pulse while the second and third terms, ``wave" like, corresponds to the dynamical modification of the interference pattern discussed above which originates from the difference of phase between the front and the back of the pulse. 
This interference effect dominates for a resonant Fabry-Perot in the tunneling regime ($D_A ,  D_B\ll 1$) where the ‘‘particle" term vanishes and one observes a purely oscillating signal $n_t=[\sin (2\pi \bar n )]/(2\pi)$,
see the right panel of Fig.~\ref{fig:gp_sine}. 
In particular for $\bar n=3/4$, one finds a {\it negative} transmitted charge $n_t=-1/(2\pi)$ which is a pure interference effect: the $e^{i 3\pi/2}$ phase of the pulse dynamically brings the Fabry-Perot cavity out of resonance and as a result, the particles coming from the left are temporarily blocked. The electrons coming from the right, on the contrary, are not affected by the pulse. Therefore the current compensation between left and right is temporarily withstood and one observes a negative net current (see the purple line in  Fig.~\ref{fig:xt-cards_currents}b for instance).

\bigskip
\noindent \tb{Discussion}\newline
The requirements to observe the above predictions experimentally are threefold.
(i) One needs a device where Fabry-Perot interferences can be observed at DC which implies
that the temperature $k_B T$ is smaller than the mean level spacing $\delta=h/2\tau_F$ of the cavity.
(ii) One needs values of $\tau_F$ long enough compared with the speed of available
pulse generators. 
(iii) An important ingredient of the modeling is that the voltage drop needs to be spatially abrupt
(with respect to the distance $L$ between the two barriers A and B). The spatial shape of the voltage drop is
controlled by the ratio between the electric $C$ and quantum $e^2\rho$ capacitances of the system, as discussed in section 8.4
of ref.~\cite{Twave_formalism}. In order to obtain a large ratio $C/(e^2\rho)$ one needs a very small density of state $\rho$ and/or to use nearby metallic gates in order to obtain an efficient screening of the charges inside the device.
Requirement (iii) requires some care but various strategies can be used to enforce it, such as depositing screening
gates close to the electron gas or using systems with extremely low density of states.
One needs $\delta \ge 10 k_B T$ in order
to fulfill (i) with a good contrast which translates into $\tau_F\le 250 ps$ for a typical dilution fridge temperature of $10 mK$. This in turn imposes a pulse duration $\tau_P \approx 100 ps$ to enter the regime of fast pulses. Such
requirements are stringent but definitely within grasp of current technology. 

There are many possible systems where the above physics could be measured. Recent progress on Thz detection were made
with carbon nanotubes\cite{McEuen2008}, for instance, although these objects are rather small (which implies small time of flight hence the THz physics). In the rest of this article, we explore an implementation, perhaps the simplest one,  where the
interferometer is constructed out of the edge states of a two dimensional electron gas
in the quantum Hall regime\cite{FabryPerot_pioneer}. The one dimensional edge states have very low density of states and can be further screened by nearby metallic gates or other nearby edge states (at filling factor two). With drift velocities
$v_D\approx 10^4-10^5 m.s^{-1}$ and a phase coherence length\cite{20mum_20mK} $L_\phi\approx 20
\mu m$ at 20mK, one finds that a rather large system of length  of a few $\mu m$ should meet the requirements.   

\begin{figure}
    \includegraphics[width=0.48\textwidth]{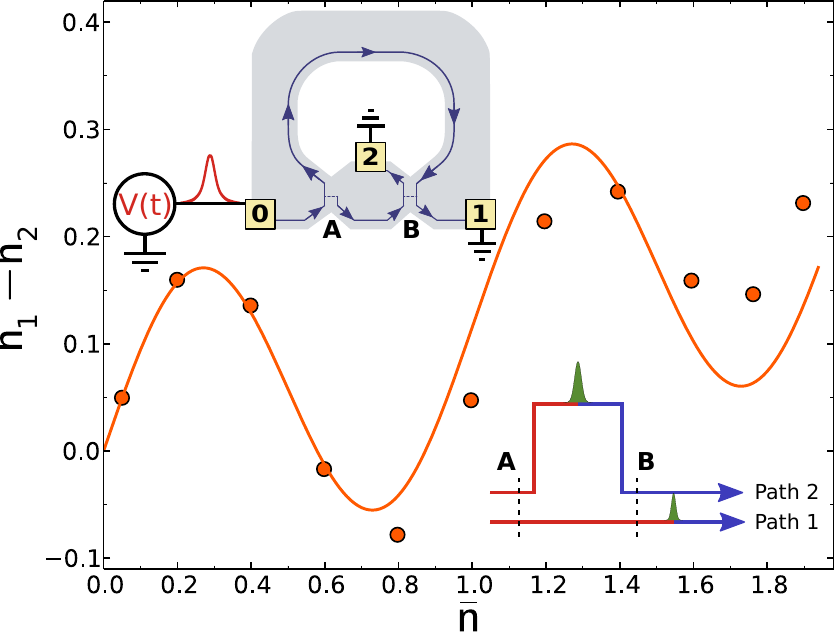}
    \caption{\tb{Voltage pulse in a Mach-Zehnder interferometer in the quantum Hall regime.} Main figure: difference $n_1-n_2$ between the transmitted charge into contact one and two as a function of the total injected charge $\bar n$. The full line corresponds
    to the analytical calculation $n_1 - n_2 = 0.12 \bar n +0.14 \sin(2 \pi \bar
    n)$ (see Methods section).  Upper inset: schematic of the system with the
    electron gas (light gray), the three contacts 0, 1, and 2 (yellow), the two
    semi-transparent quantum point contact A and B and the effective chiral edge
    states (blue arrows). Lower inset: schematic of the two paths which contribute to the 
    stationary wave function. As the pulse propagates along the different
    trajectories, a phase difference $2\pi\bar n$ appears between the front (blue) and the rear
    (red) of the pulse.} \label{fig:mz} 
\end{figure}

We simulated an electronic analogue of a Mach-Zehnder interferometer as sketched in the inset of Fig.~\ref{fig:mz}.
The device is close to the ones measured experimentally e.g. in
ref.~\cite{20mum_20mK} (although smaller due to computational limitations) and simulated in DC in ref.~\cite{Knit}.
It consists of a two dimensional gas under magnetic field with three terminals and
two quantum point contacts which serve as beam splitters. This device differs from the Fabry-Perot in two ways: first it is simpler conceptually as only two paths contribute to the transport. Second, these two paths can be resolved spatially (the edge states being chiral, transmitted and reflected waves propagate on different edge states).

Fig.~\ref{fig:mz} shows the result of the simulation for a $2\mu m^2$ sample with a density of $n_s=10^{11} cm^{-2}$, mobility
$\mu = 2\times 10^6 cm^2.V^{-1}.s^{-1}$ under a magnetic field $B=1.8T$. The
velocity is measured to be $v= 7\times 10^4 m.s^{-1}$ with an abrupt confinement of the electrons so that the difference of time of flight between the two paths is $\tau_F=64 ps$. Fast pulses of duration $\tau_P=12 ps$ were applied to electrode 0 to obtain the fast pulse limit. The system was discretized on a $3 nm$ mesh so that around $10^5$ sites were used in the simulation. The results of Fig.~\ref{fig:mz} confirm the oscillations of the transmitted charge with $\bar n$: the dynamical control of the phase between the two arms of the interferometer stands in this experimentally accessible geometry.

\begin{figure}
    \includegraphics[width=0.48\textwidth]{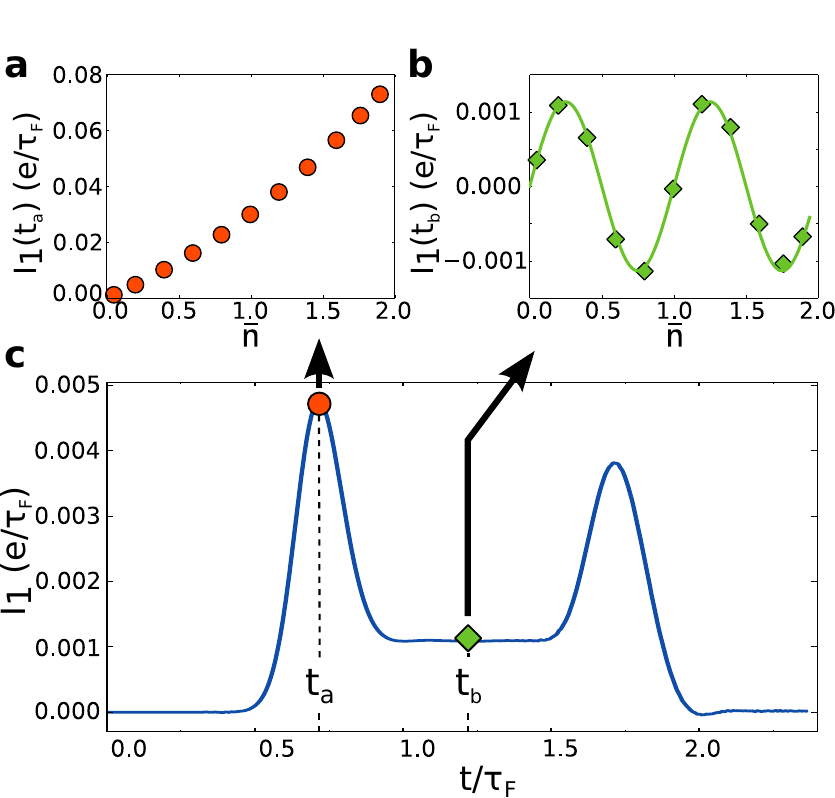}
    \caption{\label{between_peaks}\tb{Current $I_1$ at contact $1$ for the Mach-Zehnder interferometer}  (\tb{a}) and (\tb{b}) Amplitude of $I_1(t_a)$, $I_1(t_b)$, as a
    function of the number of injected particles $\bar n$. Symbols are numerical data.
    The line in (\tb{b}) corresponds to $I_1(t_b)=0.001\sin(2\pi\bar n)$. (\tb{c}) Transmitted current $I_1(t)$ as a function of time for $\bar n =0.2$. }
\end{figure}

Fig.~\ref{between_peaks}c shows the current arriving in the electrode 1 as a function of time, in direct analogy
with Fig.~\ref{fig:xt-cards_currents}b: the two peaks correspond respectively to the arrival of the pulse from the lower
arm and upper arm of the interferometer while the plateau in between corresponds to the dynamical control of the interference pattern. We show for completeness the actual value of these currents at the first peak
($t=t_a$) and on the plateau ($t=t_b$) in Fig.~\ref{between_peaks}a and Fig.~\ref{between_peaks}b respectively.
We find, as expected, that the first contribution increases with $\bar n$ while the latter oscillates as $\sin (2\pi \bar n)$. The lower inset of Fig.~\ref{fig:mz} contains a schematic of a snapshot of the interference pattern at $t=t_b$.

Experiments have now reached the technological threshold where they can be made fast and cold
enough for the physics of this proposal to be accessible in the lab\cite{Glattli}. As the
available measuring apparatus are getting faster every day, many other
nanoelectronics systems will reveal intriguing new physical effects
when probed with very fast pulses. In particular,
the concept of dynamical control of the interference pattern developed here is very generic and could be extended beyond the physics
of electronic interferometers. For instance, Andreev resonant states which form on the boundary of superconductors or
the oscillatory magnetic exchange interaction in magnetic multilayers are closely related to the Fabry-Perot physics discussed here, and could be addressed in a similar way.
The capability to simulate such time-resolved quantum nanoelectronic circuits,
demonstrated here, should play a key role in proposing and analyzing these upcoming experiments.

\bigskip
\noindent \tb{Methods}\newline
\noindent\tb{Model for the Fabry-Perot geometry.} We model the Fabry-Perot
cavity with a one dimensional Hamiltonian 
\begin{align}
\mathrm{\hat{\textbf{H}}}(t) &=\int dx
-\frac{\hbar^2}{2m}\psi^{\dagger}(x) \Delta \psi(x) \nonumber\\ &+ \epsilon(x)
\psi^{\dagger}(x) \psi(x) +\theta(-x) eV(t) \psi^{\dagger}(x) \psi(x)
\end{align} where the field operator $\psi (x)$ [$\psi^\dagger(x)$] destroys
(creates) an electron at position $x$, $V(t)$ is the voltage pulse applied on
the left electrode ($x<0$) and $\epsilon (x)$ the static potential that defines
the Fabry-Perot (for $x>0$). We discretize the model on a lattice with lattice
distance $a$ and get, 
\begin{align} 
    \mathrm{\hat{\textbf{H}}}(t) =&2\gamma  + \sum_{i=1}^{N+1} \epsilon_i c^{\dagger}_{i}c_{i}
    -\gamma \sum_{i=-\infty}^{+\infty}
    c^{\dagger}_{i+1}c_{i} \nonumber\\ & -\gamma  [e^{i\phi (t)} - 1] c^{\dagger}_{1}c_{0}
    +h.c.  
\end{align} 
where $\gamma=\hbar^2/(2ma^2)$ and $\phi (t) =
\int_{-\infty}^t dt'\ eV(t')/\hbar$ (a standard gauge transformation has
been applied to transform the time dependent potential for $i\le 0$ into
a time dependent hopping between sites $0$ and $1$). The operator $c_i$
($c^\dagger_i$) destroys (creates) an electron on site $i$. $\epsilon_i$
defines the Fabry-Perot cavity of size $L= N a$: $\epsilon_1=V_A$,
$\epsilon_{N+1}=V_B$ and $\epsilon_i=-V_0+V_g$ in the central region
$i=\in\{2,3,\dots N\}$. We use Gaussian voltage pulses of width $\tau_P$ and
maximum voltage $V_P$, 
\be 
    V(t)=V_P \exp\left(-4\log(2)\frac{t^2}{\tau_P^2}\right)
    \label{pulse}
\ee 
for which $\bar n = \kappa e V_P\tau_P  /\hbar$ where $\kappa = 1/4\sqrt{\pi\log(2)}
\approx 0.17$. Note that contrary to the noise properties\cite{Levitov1996, Lorentzian_pulses,Ivanov}, 
the total number of transmitted electrons is to a wide extent insensitive to the precise shape of the pulse.  

\bigskip
\noindent\tb{Numerical method.} The DC numerical simulations were performed with
the Kwant software package\cite{kwant}. The time-dependent simulations were
performed using the WF-D method of ref.~\onlinecite{Twave_formalism} which is summarized below.
The method consists of three steps. First, we start by solving the stationary
problem for times before the pulse. We obtain the two scattering states
$\Psi_{\alpha E}^{st}(i)$ of the system for electrons coming from the left
$\alpha=L$ and from the right $\alpha=R$ for incident energy $E$. Solving the
scattering states of a time independent Hamiltonian is a well studied problem
for which efficient techniques have been developed\cite{Wimmer_thesis}. Second, once the
pulse starts (say at $t_0<0$), we simply integrate the time dependent
Schrodinger equation with $\Psi_{\alpha E}(i,t_0)=\Psi_{\alpha E}^{st}(i)$. More
specifically, we introduce the deviation ${\bar \Psi}(i,t)$ from the stationary
wave function $\Psi_{\alpha E}(i,t) = {\bar \Psi}(i,t) + e^{-iEt} \Psi_{\alpha
E}^{st}(i)$, which satisfies $ {\bar \Psi}(i,t_0)=0$. The finite system of $N$
 sites is then embedded into a larger finite system of $2M$ sites (with typically
$M=1.5 N$)  and ${\bar \Psi}(t)$ satisfies, 
\begin{align} 
\label{eq:wbl} 
&i \partial_{t} \bar\Psi(i,t) = -\gamma [\bar\Psi(i+1,t) + \bar\Psi(i-1,t)] + (2\gamma+\epsilon_i) \bar\Psi(i,t) \nonumber\\ 
&-\gamma  \delta_{i,0} [e^{i\phi (t)} -1][e^{-iEt}\Psi_{\alpha E}^{st}(1)+\bar\Psi(1,t)]\nonumber \\ 
&-\gamma \delta_{i,1} [e^{-i\phi (t)} - 1][e^{-iEt}\Psi_{\alpha E}^{st}(0)+\bar\Psi(0,t)]\nonumber \\ 
& + (\delta_{i,-M} +\delta_{i,+M}) \Sigma (E) \bar\Psi(i,t) 
\end{align} 
where the non-Hermitian term
$\Sigma (E)= (E-2\gamma)/2\gamma^2 -i\sqrt{E/\gamma^3-E^2/4\gamma^4}$  is the so-called self-energy
of the wire.  The integration of Eq.~(\ref{eq:wbl}) for
different energies is done in parallel on
different processors using a 3$^{rd}$ order Adams-Bashforth scheme. In
the last step, the results are integrated over the different
energies to obtain the observables. In particular the  current at site
$i$ reads, 
\be 
I(i,t) = 2 \frac{e\gamma}{\hbar}  {\rm Im}  \sum_\alpha \int
\frac{dE}{2\pi} f(E)  \Psi_{\alpha E}^*(i,t) \Psi_{\alpha E}(i+1,t)
\label{eq:i} 
\ee 
where $f(E)$ is the Fermi function at temperature $T$
for the Fermi energy $E_F$. This method allows one to study systems with
$10^4-10^5$ sites on a simple cluster and is expected to scale beyond $10^6$
sites on a supercomputer.

\begin{figure}
    \includegraphics[width=0.47\textwidth]{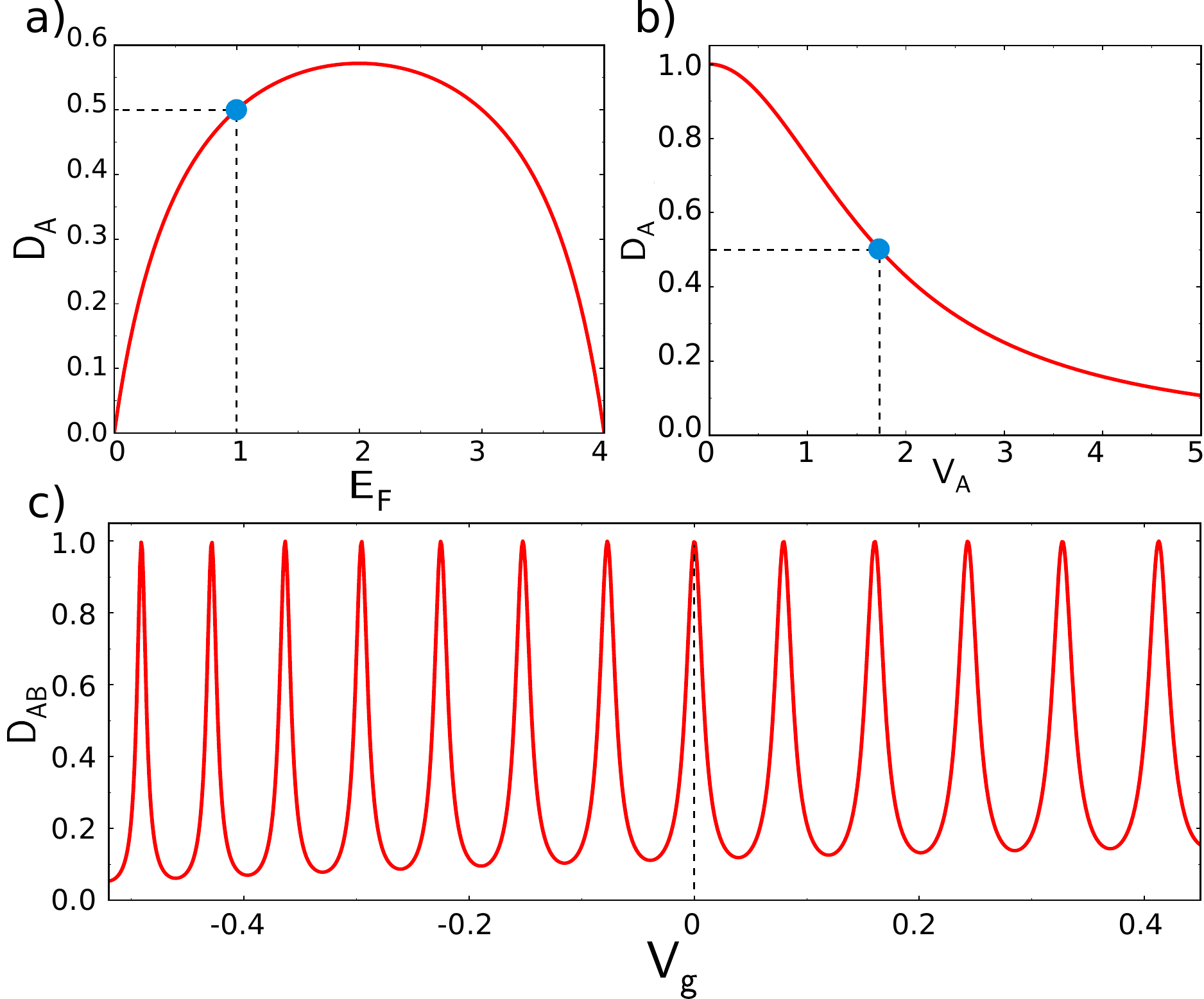}
    \caption{\tb{DC characterization of the Fabry-Perot cavity.}
    (\tb{a}) Transmission of the barrier A as a function of energy $E_F$ with $V_{A}=1.73\gamma$. 
    (\tb{b}) Transmission of barrier A as a function of
$V_A$ for $E_F=1\gamma$. (\tb{c}) Transmission probability $D_{AB}$ of the entire system with two barriers as a function of $V_g$ 
for $V_A=V_B=1.73\gamma$ and $E_F=1\gamma$ ($D_A=D_B=0.5$). The parameters of panel (\tb{c}) correspond to the blue circles of panel (\tb{a}) and (\tb{b}).   \label{fig:DC_charac}}
\end{figure}

\bigskip
\noindent\tb{Parameter set for the Fabry-Perot geometry.} Most of the data presented here were obtained with the
following set of parameters: $N=70$ and $E_F=\gamma$ so that $\tau_F\approx 35\gamma^{-1}$ and $\delta\approx 0.09\gamma$. Various durations of the pulses were used
from $\tau_P=5\gamma^{-1}$ to $\tau_P=100\gamma^{-1}$. We found that $\tau_P\ge 5\gamma^{-1}$ is necessary to enforce $\hbar/\tau_P \ll E_F$
and get rid of spurious effects associated with the band width of the model.
The values of $V_A$ and $V_B$ are given in Fig.~\ref{fig:DC_charac}b while
$V_0=-1.068$.

\bigskip
\noindent\tb{A comment on electron-electron interactions.} A common difficulty
encountered in time dependent transport, which was pointed out by Buttiker some
years ago\cite{Buttiker_dynamic_conductance}, is the crucial role of electrostatics in
restoring a gauge invariant, current-conserving theory. Indeed, in the
non-interacting theory used here, the conservation equation for the charge
reads, \be \partial_t \rho(x,t) + \partial_x I(x,t)=0 \ee where $\rho(x,t)$ is
the charge density and $I(x,t)$ the local current. In presence of time dependent
perturbations (such as the voltage pulse), the current is not conserved and a
finite charge density {\it temporary} accumulates in the system. An accumulation
of charge costs however a tremendous amount of electrostatic energy so that in
real systems, this charge density is screened by image charges in nearby gates.
Those image charges  result in a displacement current $I_d= \partial_t
\rho(x,t)$ flowing in those electrodes. Only once this displacement current is
taken into account does one recover current conservation. As a result of the
presence of this time dependent charge density, one should, {\it at the mean
field level} include the corresponding time dependent potential created by these
charges into our time dependent Schrodinger equation. Let us make four specific
remarks for the situation studied in this article.  First, we study situations with
a small number of injected particles $\bar n$, therefore one should be very
careful with the mean field approach as one wants to avoid spurious self
interacting terms present at the Hartree level. Second, all our calculations are
done for a non-interacting model, and are therefore {\it a priori} expected to
be valid in presence of metallic gates in close proximity to the quantum wire.
Third, while the displacement currents and corresponding time dependent potentials
can modify the a.c. properties of the system, the total transmitted charge
$n_t$ shall not be affected by treating explicitly the electrostatic problem.
Indeed, the total number of transmitted and reflected electrons form  {\it
conserved and gauge invariant quantities} (in the sense defined by Buttiker\cite{Buttiker_dynamic_conductance}) and
therefore do not suffer from the flaws of their a.c. counterparts. 
In plainer words, the integral (over time) of the displacement
currents as well as the corresponding time dependent potentials is zero,
therefore their presence do not modify $n_t$. Finally, recent
experiments\cite{Glattli} with fast voltage pulses
indicate that the non-interacting theory works remarkably
well for those systems. A longer discussion of current conservation and gauge
invariance can be found in ref.~\onlinecite{Twave_formalism}. 

Note that beside the above mentioned aspects, the electrostatics remains crucial in the determination of
the spatial profile of the voltage drop created by the voltage pulse. In order to observe the effects discussed in
this article, one needs to be able to create spatially localized voltage drops
that can subsequently propagate inside the interferometer. The corresponding
condition has been discussed in section 8.4 of ref.~\onlinecite{Twave_formalism}
which we refer to.

\bigskip

\noindent\tb{Calibration of the Fabry-Perot geometry.}
Fig.~\ref{fig:DC_charac} shows the DC characteristics which were used to calibrate our device.
Fig.~\ref{fig:DC_charac}a and Fig.~\ref{fig:DC_charac}b show the transmission probability of
a single barrier, say A,  as a function of the Fermi energy (a) and $V_A$. 
Fig.~\ref{fig:DC_charac}c shows the transmission probability (conductance in unit of $e^2/h$)
of the full Fabry-Perot cavity as a function of the gate voltage $V_g$
from which we can extract the peak to peak mean level spacing $\delta=0.09\gamma$.
 
\begin{figure}
    \includegraphics[width=0.47\textwidth]{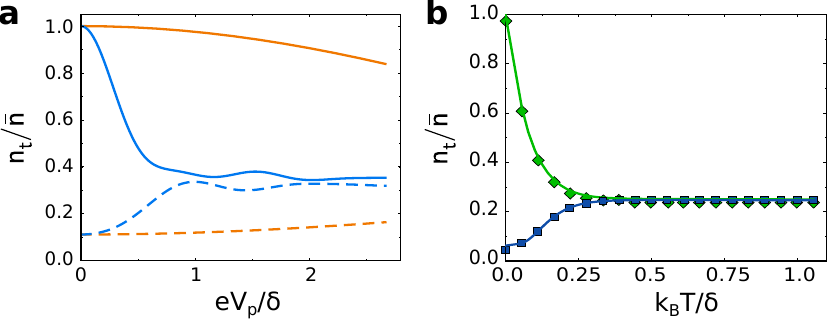}
    \caption{\tb{Effect of voltage amplitude and temperature on the visibility
of the interference pattern.} (\tb{a}) Transmission probability $n_t/\bar n$ as
a function of $V_P/\delta$ for a system at resonance (full lines, $V_g=0$) and
off resonance (dashed line, $V_g=\delta/2$) for a short (orange,
$\tau_P=\tau_F/7$) and long (blue, $\tau_P=3\tau_F$) pulse. $D_A=D_B=0.5$.
(\tb{b}) Transmission probability $n_t/\bar n$ as a function of temperature for
the same short pulse and $V_P=0.5\delta$. Symbols: numerical results,  lines:
energy  average $\langle -D_{AB}(V_g,E)\partial_E f(E)\rangle_E$. The upper
curves correspond to $V_g=0$ (resonance) while the lower one is off resonance
$V_g=\delta/2$.\label{fig: other_nt} } 
\end{figure}

Fig.~\ref{fig: other_nt}a shows the resonant and off resonance signal $n_t/\bar n$ as a function of the maximum voltage $V_P/\delta$ for both the short and long pulses.
As the visibility of the fast pulses is sensitive to $\bar n$ and not to $V_P/\delta$ [Eq.~(\ref{wp})], we find that the system can retain a high visibility for $V_P>\delta$ while
the interference pattern of the long pulse is totally smeared out.
We study in Fig.~\ref{fig: other_nt}b the temperature dependence of $n_t$ at and off resonance. We find that a low
$k_BT\le 0.1\delta$ temperature is needed to observe interferences with a good visibility. This requirement is as stringent as the
DC requirement but not more, so that temperature should not be a restriction
for the observation of the effects predicted in this work.

\begin{figure}
    \includegraphics[width=0.47\textwidth]{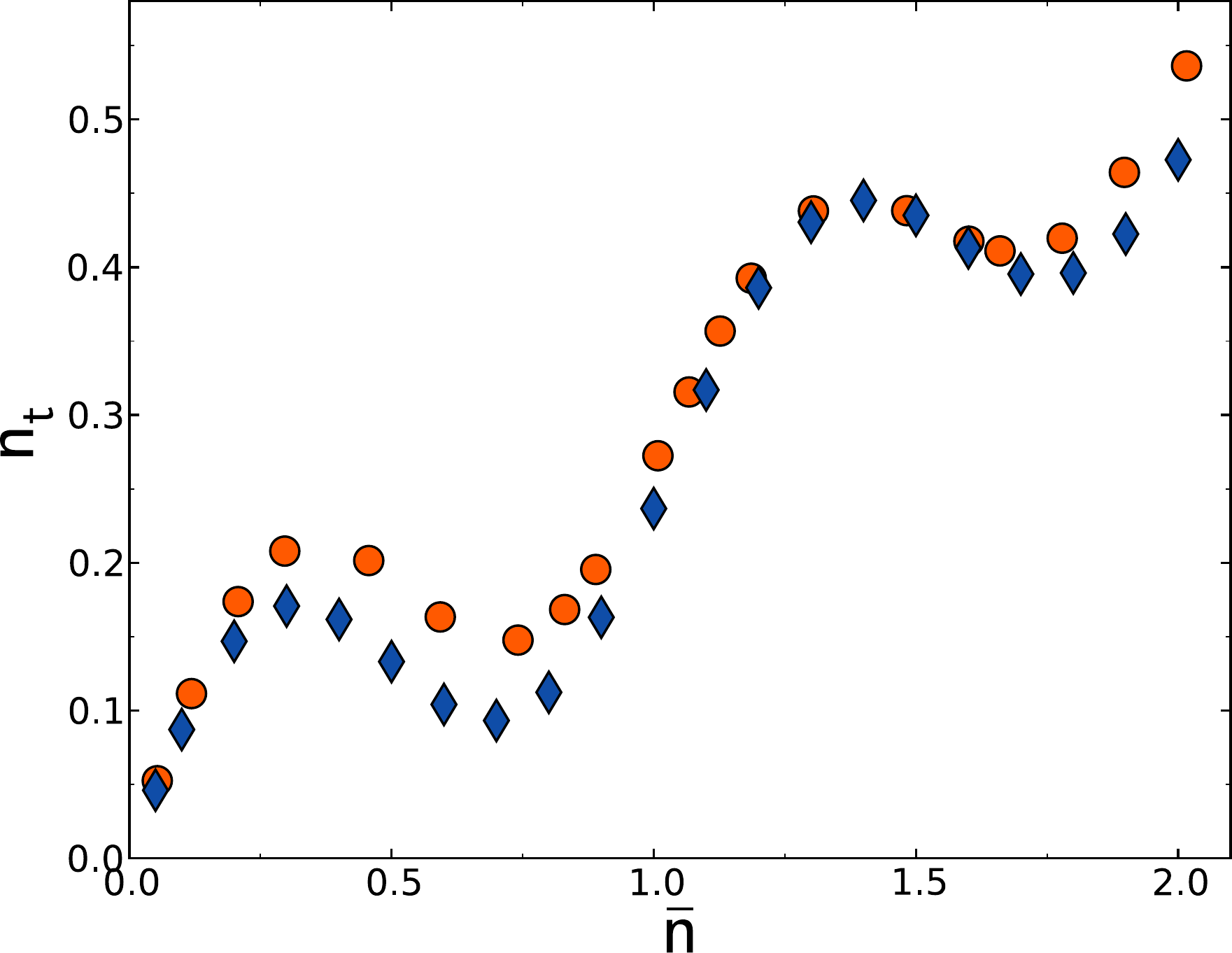}
    \caption{\label{lorentzian}\tb{Effect of the shape of the pulse.} Transmitted charge $n_t$ as a function of total
    injected charge $\bar n$. The system is at resonance and $V_P$ is varied
    with $D_A=D_B=0.5$. Orange circles are the data of the short Gaussian
    pulse case from Fig.~\ref{fig:gp_sine}b, blue diamonds correspond to a
    Lorentzian shaped pulse with width $\tau_P=\tau_F/7$.}
\end{figure}

Last, Fig.~\ref{lorentzian} presents the number of transmitted electrons as a function of the injected one
for two different pulse shapes: a Gaussian pulse [Eq.~(\ref{pulse})] and a Lorentzian one
($V(t) = V_P / (1 + 4t^2/\tau_P^2)$). We find, as expected from the analytical
calculation, that the results are insensitive to the shape of the pulse in the
fast pulse limit and we recover the oscillating behavior with respect to $\bar
n$. We emphasize that this is in sharp contrast with the current noise in the
single barrier case studied in ref.~\onlinecite{Lorentzian_pulses}

\bigskip
\noindent\tb{Analytical technique for the calculation of $n_t$.}
Our starting point for the calculation of $n_t$ is Eq.~(95) of ref.~\onlinecite{Twave_formalism},  
\begin{align}
    \label{srt} 
    n_{t} = &\int \frac{dE}{2\pi}\frac{dE'}{2\pi} |d(E',E)|^2
    [f(E)-f(E')] 
\end{align} 
where $d(E',E)$ is the amplitude of probability for
an incident electron coming from the left with energy E to be transmitted
with energy $E'$. $d(E',E)$ can be further decomposed into 
\be
    d(E',E)=d_v(E'-E)d_{AB}(E') 
\ee  
where the first (inelastic) term originates
from the voltage drop while the second comes from the (elastic) Fabry-Perot
cavity.  In order to derive Eq.~(\ref{srt}), we have made use of the fact
that the transmission amplitude $d_v'(E'-E)$ for electrons coming from the
right is given by $d_v'(E'-E)=d_v^*(E-E')$. Note that Eq.~(\ref{srt}) as a
whole is a perfectly convergent integral whose integrand is concentrated
around the Fermi level (assuming the voltage pulse is slow enough compared
with $\hbar/E_F$). However each of its two sub terms spread over the entire
band of the model, so one should refrain from calculating these two terms
separately, if possible. Eq.~(\ref{srt}) has a nice straightforward
interpretation: one simply sums over the (incoherent) incoming states and
calculate their total transmission probabilities regardless of the final
energy. In the absence of voltage pulse the vanishing $n_t$ comes from the
compensation between electrons coming from the left and from the right.

Our model for the Fabry-Perot transmission amplitude has been given in the core
of the text. To calculate $d_v(E'-E)$, one defines the scattering states on both
sides of the voltage drop: 
\begin{align} 
    \Psi_L(n,t)&=\frac{1}{\sqrt{v(E)}}
e^{ik(E) n -iEt}  \\ &+ \int \frac{dE'}{2\pi} r_v(E',E) \frac{1}{\sqrt{v(E')}}
e^{-ik(E') n -iE't}\nonumber 
\end{align} 
on the left and 
\be 
\Psi_R(n,t)= \int \frac{dE'}{2\pi} d_v(E',E) \frac{1}{\sqrt{v(E')}} e^{ik(E') n -iE't} 
\ee 
on the
right (with $v(E)=\partial E/\partial k$ the velocity associated with the
dispersion relation $E=2\gamma \cos k$). By ‘‘matching" the left and right waves
across the voltage drop (see ref.~\onlinecite{Twave_formalism}), one obtains a set of equations for
$d_v$ and $r_v$. In the limit where the pulse is slow $\tau_P \gg
\hbar/E_F$, and $V_P\ll E_F/e$ is low compared with the Fermi energy, (the case of interest for our
nanodevices), we can linearize the dispersion relation and we simply recover the
result of ref.~\onlinecite{Lorentzian_pulses}, 
\be 
\label{dd} 
    d_v(E'-E)=\int dt e^{i (E'-E)t} e^{-i\phi(t)} 
\ee 
with $r_v=0$. To proceed, we expand $d_{AB}(E)$ in terms of the
different paths, 
\be 
d_{AB}(E)=\sum_{n=0}^{\infty} d_A d_B (r_A r_B)^n e^{2i\tau_F (E+eV_g) n} 
\ee 
and introducing $\epsilon=E'-E$, we get, 
\begin{align}
    n_{t} = &\int \frac{dE}{2\pi}\frac{d\epsilon}{2\pi}\sum_{n,m}
    |d_v(\epsilon)|^2 D_A D_B (r_A r_B)^{n+m} \nonumber\\ &\times e^{2i\tau_F
(E+eV_g) (n-m)}[f(E-\epsilon)-f(E)] 
\end{align} 
We can now perform the
integration over $E$ (at zero temperature) which binds together the two parts of
the integral.  The terms $n=m$ and $n\ne m$ need to be considered separately,
and we get, 
\begin{align} 
    n_{t}& = D_{AB}^{cl}\int \frac{d\epsilon}{2\pi}\
|d_{v}(\epsilon)|^2 \epsilon + \int \frac{d\epsilon}{2\pi}\  |d_{v}(\epsilon)|^2
\frac{D_{A}D_{B}}{2\pi} \nonumber\\ &\times \sum_{n\ne m} (r_{A}r_{B})^{n+m}
\frac{e^{i\alpha_{g}(n-m)}}{i2\tau_{F}(n-m)} (e^{i2\tau_{F}\epsilon(n-m)} - 1)
\end{align} 
with $\alpha_{g} = 2\tau_{F}(E_{F} + eV_{g})/\hbar$. We can now
replace $d_v(\epsilon)$ by its expression Eq.~(\ref{dd}) and performing the integral
over $\epsilon$, we arrive at
\begin{align}
\label{eq:general}
    n_t = D_{AB}^{cl} \bar n + &\sum_{n} \sum_{m\neq n} \frac{D_A D_B}{2\pi} (r_A
    r_B)^{n+m} \frac{e^{i\alpha_g (n-m)}}{i 2\tau_F(n-m)}  \\ \nonumber
    &\times\int dt\ \left[e^{-i\phi(t)}e^{i\phi(t+2\tau_F(n-m))} -1\right]
\end{align}
Eq.~(\ref{eq:general}) applies for all pulses, short and long. Assuming an infinitely short pulse $\phi (t)=\theta (t) e^{i2\pi\bar n}$, we obtain after integration and
resummation of the geometric series,
\begin{align} 
    n_{t}\big|_{short} &= D_{AB}^{cl}\ \bar{n} + (D_{AB}(V_g) - D_{AB}^{cl})
\frac{\sin(2\pi\bar{n})}{2\pi}\nonumber\\ &-
\frac{2D_{AB}(V_g)D_{AB}^{cl}r_{A}r_{B}}{\pi D_{A}D_{B}} \sin^{2}(\pi \bar{n})
\sin(2\pi V_g/\delta) 
\end{align}
In the case of very long pulses $\phi(t)$ evolves
very slowly with respect to $\tau_F$ so that one expands $\phi (t+ a
\tau_F)\approx \phi(t) +a \tau_F eV(t)/\hbar$. In this limit,
Eq.~(\ref{eq:general}) allows one to recover the adiabatic 
result,
\be
n_t\big|_{long} = \int dt \int_{E_F}^{E_F+V(t)} \frac{dE}{2\pi} D_{AB}(E)
\ee

The calculation for the Mach-Zehnder geometry proceeds along the same lines, and is even simplified by the presence of only two paths contributing to the transmission amplitude of the device. The transmission probabilities from lead $0$ to $1$ ($2$) reads,
\begin{align}
    |S^0_{10}(E)|^2 &= D_AD_B + R_AR_B  \\ &+ 2\sqrt{D_A D_B R_A R_B}\cos(\phi + \tau_F(E-E_F)),\nonumber
\end{align}
\begin{align}
    |S^0_{20}(E)|^2 &= D_AR_B + R_AD_B  \\ &- 2\sqrt{D_A D_B R_A R_B}\cos(\phi + \tau_F(E-E_F)),\nonumber
\end{align}
with $\phi$ the total magnetic flux through the central depleted region (in unit of $\hbar/e$) and $\tau_F$ the extra time needed for the upper paths with respect to the lower one. After following the same steps as for the Fabry-Perot geometry, one obtains (in the limit of short pulses) the number of particles transmitted to contact $1$ ($2$),
\begin{align}
\label{n1}
    n_{1} &= (D_AD_B + R_AR_B) \bar n  \\ &+ \frac{2}{\pi}\sqrt{D_A D_B R_A R_B} \sin(\pi
    \bar n)\cos(\pi \bar n + \phi)\nonumber
\end{align}
\begin{align}
\label{n2}
    n_{2} &= (D_AR_B + R_AD_B) \bar n \\ 
                 &-\frac{2}{\pi}\sqrt{D_A D_B R_A R_B} 
                \sin(\pi \bar n)\cos(\pi \bar n + \phi)\nonumber
\end{align}

\bigskip
\noindent\tb{Model for the Mach-Zehnder geometry.}
\begin{figure}[t!]
    \includegraphics[width=0.45\textwidth]{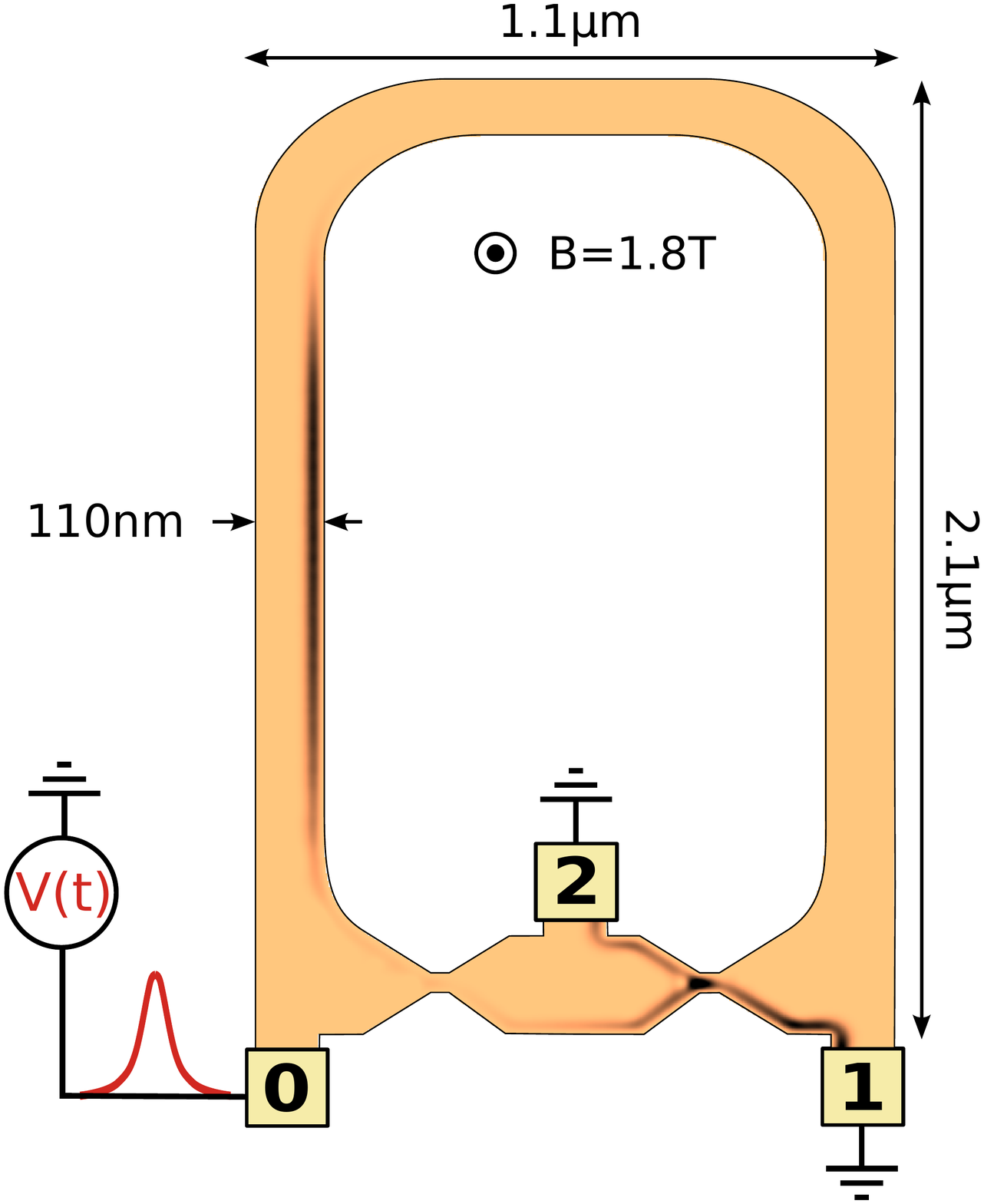}
    \caption{\label{mz-device} \tb{Mach-Zehnder interferometer}. Snapshot
            of the local electronic density at $t=46ps$. The color map
            indicates the deviation from equilibrium which goes from 0 (salmon) to $0.22\times
            10^{11}cm^{-2}$ (black).} 
\end{figure}
We consider a two-dimensional electron gas made from a two-dimensional
GaAs/AlGaAs heterostructure with high mobility $\mu= 2\times 10^6
cm^2.V^{-1}.s^{-1}$, an electronic density $n_s=10^{11}cm^{-2}$ and a
perpendicular magnetic field $B=1.8T$ (corresponding to filling factor one,
first Hall plateau). The three terminal electronic Mach-Zehnder interferometer
is sketched in Fig.~\ref{mz-device}.

The system is modeled within the effective mass approximation in presence
of a small static disorder. The Schrodinger equation is discretized on a mesh with a step
$a= 3 nm$ much smaller than both the Fermi wave length $\lambda_F= 79 nm$ and magnetic length $l_B= 19 nm$ of the system.
The magnetic field is accounted through a standard Peierls' substitution. The model and its DC characterization (with a slightly different geometry) were discussed in ref.~\onlinecite{Knit}.

In the simulations, contacts $1$ and $2$ are grounded,
while a voltage pulse is applied on contact $0$ [same pulse as Eq.~(\ref{pulse})].
The injected current follows the edge state and is split into two parts as it
reaches the first quantum point contact (QPC). Both QPCs are set to be
semi-transparent $D_A=D_B=0.5$ and consequently act as beam splitters. The two parts of
the initial current are recombined at the second QPC. Fig.~\ref{mz-device} actually corresponds to a snapshot
of the simulation at an intermediate time $t=46 ps$: the color code indicates the deviation of the local electronic density with respect to the equilibrium value. At this intermediate time, the pulse has already passed through the first QPC and is split into two parts. The lower (transmitted) part is reaching the electrodes 1 and 2 while the upper (reflected) part is traveling along the longer arm of the interferometer.


\bigskip
\noindent\tb{References} \newline

\bigskip
\noindent\tb{Acknowledgment} \newline 
This work was supported by the ERC grant MesoQMC from the
European Union.  We thank C. Groth, P. Roche, M. Houzet, D.C. Glattli, C. Bauerle, P. Roulleau and D. Luc for very
valuable discussions.

\bigskip
\noindent\tb{Author contributions}\newline
X.W. initiated the project. B.G. performed the numerical simulations. X.W. and B.G. performed the analytical calculations, data analysis and wrote the manuscript.

\bigskip
\noindent\tb{Contact information}\newline
Correspondence and requests for materials should be addressed to X.W. (email: xavier.waintal@cea.fr)

\end{document}